\documentstyle{article}
\begin{document}
\newcommand{\dfrac}[2]{\frac{\displaystyle #1}{\displaystyle #2}}
\baselineskip=12pt
\title{\bf
Propagators for scalar bound states at finite temperature in a NJL model  \thanks{ This work was partially supported by the National Natural Science Foundation of China and by Grant No.LWTZ-1298 of the Chinese Academy of Sciences.}
 \\}
\author{
{\bf
Bang-Rong Zhou
} \\
\normalsize Department of Physics, The Graduate School of The Chinese Academy of Sciences \\
\normalsize Beijing 100039, China \\
\normalsize  and CCAST (World Laboratory) P.O. Box 8730, Beijing 100080, China \\
}
\date{}
\maketitle 
\begin{abstract}
We reexamine physical causal propagators for scalar and pseudoscalar bound states 
at finite temperature in a chiral $U_L(1)\times U_R(1)$ NJL model, defined by 
four-point amputated functions subtracted through the gap equation, and prove that 
they are completely equivalent in the imaginary-time and real-time formalism by 
separating carefully the imaginary part of the zero-temperature loop integral. It 
is shown that the thermal transformation matrix of the matrix propagators for these 
bound states in the real-time formalism is precisely the one of the matrix propagator 
for an elementary scalar particle and this fact shows similarity of thermodynamic 
property between a composite and an elementary scalar particle. The retarded and 
advanced propagators for these bound states are also given explicitly from the 
imaginary-time formalism.
\end{abstract}
\\
\section{Introduction}
Finite temperature field theory has attracted much research interest
of people due to its application to evolution of early universe and phase transition 
of the hadron matter [1-5]. However, the complete equivalence between its two
formalisms - the imaginary-time and the real-time formalism [4]- has been a
subtle issue. It is usually assumed that the two formalisms should give the
same results. Nevertheless, in some actual problems where only amputated
Green functions are involved, the calculations in the two formalisms often show
different results [6]. Much work has been contributed to seeking correspondence 
between the two formalisms for some amputated functions [7-10]. The general 
conclusion is that the difference between the amputated functions obtained in the 
two formalisms could originate from that one actually deal with different Green 
functions in the two cases thus they should be used for different physical purposes 
[8]. In a recent research on the Nambu-Goldstone mechanism of dynamical electroweak 
symmetry breaking at finite temperature [11] based on a Nambu-Jona-Lasinio (NJL) 
model [12], we calculate the propagators for scalar bound states which show different 
imaginary parts in their denominators in the two formalisms. It was naturally supposed 
that we had calculated different Green functions in the two cases. However, this 
inference is open to question. The reason is that the analytic continuation used there 
of the Matsubara frequency to real energy, though not in the most general form, was essentially made as the way leading to 
a causal Green function which should just be that obtained in the real-time formalism. 
In addition, one notes that the propagators for scalar bound states at finite 
temperature in a NJL model correspond some four-point amputated functions, 
and the calculations of four-point amputated functions can be effectively reduced 
to the ones of some two-point functions.  It is accepted that a two-point function 
should be equivalent in the two formalisms of thermal field theory [7]. Therefore, 
it is necessary for us to reexamine the whole calculations of a NJL model. In this 
paper, we will do that by means of a chiral $U_L(1)\times U_R(1)$ NJL model. After 
rigorous and careful calculations, we will finally prove that the propagators for 
scalar bound states including the imaginary parts in their denominators are in fact 
identical in the two formalisms. We also find a remarkable result that, in the 
real-time formalism, the thermal transformation matrix of the matrix propagators for 
scalar bound states is precisely the one of the matrix propagator for an elementary  
scalar particle. The key-points to reach the above conclusions which were ignored 
in Ref.[12] are that, except keeping the most general form of the analytic 
continuation in the imaginary-time formalism, we must carefully consider and separate 
the imaginary part of the zero-temperature loop integral from relevant expressions. 
This is very similar to the case of a complete calculation of a two-point 
function [13], except that in the case of a NJL model we must additionally use the 
gap equation so as to eliminate some parts of the Green functions. We will discuss 
the propagators for scalar bound states in the one-loop approximation respectively 
in the imaginary-time and real-time formalism , then compare the derived results.
\section{The imaginary-time formalism}
In the imaginary-time formalism, the Lagrangian of the four-fermion
interactions will be
\[
{\cal L}^I_{\rm 4F}=
\frac{G}{4}[(\bar{\psi}\psi)^2-(\bar{\psi}\gamma_5\psi)^2],
\]
where $\psi$ is the fermion field with single flavor and $N$ colors, $G$ is
the coupling constant.  Assume the scalar interactions in ${\cal L}^I_{\rm 4F}$ may
 lead to the formation of the fermion condensate ${\langle\bar{\psi}\psi\rangle}_T$ 
at temperature $T$, then we will obtain the gap equation which determines the 
dynamical fermion mass $m(T)\equiv m$ [14]
\begin{equation}
1=GI, \ \ I=2N\int\frac{d^3l}{(2\pi)^3}\,T\sum_{n=-\infty}^{\infty}
\frac{1}{(\omega_n+i\mu)^2+{\stackrel{\rightharpoonup}{l}}^2+m^2},
\end{equation}
where $\omega_n=(2n+1)\pi/\beta\, (n=0,\pm 1, \pm 2, ...; \beta=1/T)$ is the 
Matrubara frequency of the fermions and $\mu$ is the chemical potential of the 
fermions. A scalar bound state $\phi_S=(\bar{\psi}\psi)$ could be formed only through 
the four-fermion interactions, so we must define the propagators of $\phi_S$ as the 
four-point  amputated function $\Gamma_I^{\phi_S}(-i\Omega_m, \stackrel{\rightharpoonup}{p})$ for the transition from $(\bar{\psi}\psi)$ to $(\bar{\psi}\psi)$, where 
$\Omega_m=2\pi m/\beta (m=0,\pm 1, \pm 2, ...)$ is the Matsubara frequency 
corresponding to the external energy. $\Gamma_I^{\phi_S}(-i\Omega_m,
\stackrel{\rightharpoonup}{p})$ submits to the the Schwinger-Dyson equation
\begin{equation}
\Gamma_I^{\phi_S}(-i\Omega_m, \stackrel{\rightharpoonup}{p})
=\frac{G}{2}[1+2N(-i\Omega_m, \stackrel{\rightharpoonup}{p})
\Gamma_I^{\phi_S}(-i\Omega_m, \stackrel{\rightharpoonup}{p})],
\end{equation}
where the fermion loop contribution
\begin{eqnarray*}
N(-i\Omega_m, \stackrel{\rightharpoonup}{p})&=&
2N\int\frac{d^3l}{(2\pi)^3}\,T\sum_{n=-\infty}^{\infty}
\frac{(\omega_n+i\mu) (\omega_n+i\mu+\Omega_m)
+ \stackrel{\rightharpoonup}{l}\cdot
(\stackrel{\rightharpoonup}{l}+\stackrel{\rightharpoonup}{p})-m^2}
{[(\omega_n+i\mu)^2+\omega_l^2][ (\omega_n+i\mu+\Omega_m)^2+\omega_{l+p}^2]}
\\
&=&I+2N\int\frac{d^3l}{(2\pi)^3}\,T\sum_{n=-\infty}^{\infty}
\frac{-(\Omega_m^2+{\stackrel{\rightharpoonup}{p}}^2)-2m^2-
(\omega_n+i\mu)\Omega_m- \stackrel{\rightharpoonup}{l}\cdot
\stackrel{\rightharpoonup}{p}}
{[(\omega_n+i\mu)^2+\omega_l^2][ (\omega_n+\Omega_m+i\mu)^2+\omega_{l+p}^2]}
\end{eqnarray*}
with the denotations $\omega_l^2={\stackrel{\rightharpoonup}{l}}^2+m^2$ and 
$\omega_{l+p}^2=(\stackrel{\rightharpoonup}{l}+\stackrel{\rightharpoonup}{p})^2+
m^2$. By means of the gap equation (1), we can write the solution of (2) as
\[
\Gamma_I^{\phi_S}(-i\Omega_m, \stackrel{\rightharpoonup}{p})
=-1/2N[-(\Omega_m^2+{\stackrel{\rightharpoonup}{p}}^2)-4m^2]
\int\frac{d^3l}{(2\pi)^3}
A(-i\Omega_m, \stackrel{\rightharpoonup}{p}, \stackrel{\rightharpoonup}{l}),
\]
\begin{equation}
A(-i\Omega_m, \stackrel{\rightharpoonup}{p}, \stackrel{\rightharpoonup}{l})=
T\sum_{n}\frac{1}{(\omega_n+i\mu)^2+\omega_l^2}
\frac{1}{(\omega_n+\Omega_m+i\mu)^2+\omega_{l+p}^2},
\end{equation}
where we  have used the property that, owing to the Lorentz invariance,
$\Gamma_I^{\phi_S}(-i\Omega_m, \stackrel{\rightharpoonup}{p})$
must be a function of $-(\Omega_m^2+{\stackrel{\rightharpoonup}{p}}^2)$ ,
thus 
\[
\Gamma_I^{\phi_S}(-i\Omega_m, \stackrel{\rightharpoonup}{p})=
\Gamma_I^{\phi_S}(i\Omega_m, -\stackrel{\rightharpoonup}{p})=
\Gamma_I^{\phi_S}(-i\Omega_{-m}, -\stackrel{\rightharpoonup}{p}).
\]
Doing the sum of the Matsubara frequency $\omega_n$ in (3) by the standard procedure 
[11,15] then making the analytic continuation of the external energy
$-i\Omega_m \rightarrow p^0+i\varepsilon p^0 \, (\varepsilon=0_+)$ and keeping the general form of the replacement, we may write
$ A(-i\Omega_m \to p^0+i\varepsilon p^0, \stackrel{\rightharpoonup}{p}, \stackrel{\rightharpoonup}{l})$ by
\begin{eqnarray*}
A(p, \stackrel{\rightharpoonup}{l})
&=& \frac{1}{4\omega_{l}\omega_{l+p}}\left\{
\frac{1-n(\omega_{l}+\mu)-n(\omega_{l+p}-\mu)}
     {-p^0+\omega_{l}+\omega_{l+p}-i\varepsilon \eta(p^0)}
+\frac{n(\omega_{l}+\mu)-n(\omega_{l+p}+\mu)}
     {-p^0+\omega_{l}-\omega_{l+p}-i\varepsilon \eta(p^0)}\right. \\
&&\left.-\frac{n(\omega_{l}-\mu)-n(\omega_{l+p}-\mu)}
     {-p^0-\omega_{l}+\omega_{l+p}
            -i\varepsilon \eta(p^0)}
  - \frac{1-n(\omega_{l}-\mu)-n(\omega_{l+p}+\mu)}
     {-p^0-\omega_{l}-\omega_{l+p}-i\varepsilon \eta(p^0)} \right\},
\end{eqnarray*}
where $n(\omega_l\pm \mu)=1/[e^{\beta(\omega_l\pm \mu)}+1]$ and
$\eta(p^0)=p^0/|p^0|$. For the
convenience of making a comparison with the following results in the real-time
formalism we express $ A(p, \stackrel{\rightharpoonup}{l})$ as a integral of
$l^0$ by using
\begin{equation}
\sin^2\theta(l^0,\mu)=\frac{\theta(l^0)}{\exp[\beta (l^0-\mu)]+1}
                        +\frac{\theta(-l^0)}{\exp[\beta (-l^0+\mu)]+1}
\end{equation}
and the formula $1/(X+i\varepsilon)=X/(X^2+\varepsilon^2)-i\pi\delta(X)$.
Eventually we obtain the physical causal propagator for $\phi_S$ in the imaginary-time formalism
\begin{eqnarray}
\Gamma_{IF}^{\phi_S}(p)&\equiv &i\Gamma_I^{\phi_S}(-i\Omega_m \to
p^0+i\varepsilon p^0, \stackrel{\rightharpoonup}{p})\nonumber \\
&=&-i/(p^2-4m^2+i\varepsilon)[K(p)+H(p)-iS^I(p)],
\end{eqnarray}
where
\begin{eqnarray}
K(p)&=&-2N\int\frac{id^4l}{{(2\pi)}^4}\frac{1}{(l^2-m^2+i\varepsilon)
         [(l+p)^2-m^2+i\varepsilon]} \nonumber \\
      &=&\frac{N}{8\pi^2}\int \limits_{0}^{1}dx\left(
          \ln\frac{\Lambda^2+M^2}{M^2}-\frac{\Lambda^2}{\Lambda^2+M^2}
          \right), \ \ M^2=m^2-p^2x(1-x)
\end{eqnarray}
is the contribution from the zero temperature fermion loop with the four-dimension 
Euclidean momentum cut-off $\Lambda$,
\begin{equation}
H(p)=4\pi N\int \frac{d^4l}{{(2\pi)}^4}\left\{
         \frac{(l+p)^2-m^2}{{[(l+p)^2-m^2]}^2+\varepsilon^2}+(p\to -p)
         \right\}\delta(l^2-m^2)\sin^2\theta(l^0,\mu)
\end{equation}
and
\begin{eqnarray}
S^I(p)&=&\eta(p^0)4\pi^2N\int
     \frac{d^4l}{{(2\pi)}^4}\delta(l^2-m^2)\delta[(l+p)^2-m^2] \nonumber \\
   & &\times [\sin^2\theta(l^0,\mu)\eta(l^0+p^0)+\sin^2\theta(l^0+p^0, \mu)\eta(-l^0)]. 
\end{eqnarray}
It is emphasized that $S^I(p)$ does not contain any pinch singularity due to the 
factors
$\eta(l^0+p^0)$ and $\eta(-l^0)$  in its integrand and since
\[
\delta(l^2-m^2)\delta[(l+p)^2-m^2]=0, \;\; {\rm when} \;\; 0\leq p^2<4m^2,
\]
we must have $S^I(p)=0$, when $0\leq p^2<4m^2 $. In addition, from
(6), $K(p)$ is real when $p^2<4m^2$ and from (7), $H(p)$ is always real. [16]. 
Similarly, we may find out the physical causal propagator for the pseudoscalar bound 
state $\phi_P=(\bar{\psi}i\gamma_5\psi)$
\begin{equation}
\Gamma_{IF}^{\phi_P}(p)=-i/(p^2+i\varepsilon)[K(p)+H(p)-iS^I(p)].
\end{equation}
The equations (5) and (9)  show that $\phi_S$ and $\phi_P$ each have the mass $2m$ 
and $0$ 
thus can be respectively identified with the massive "Higgs" scalar particle and 
the massless Nambu-Goldstone boson for the spontaneous symmetry breaking
$U_L(1)\times U_R(1) \to U_{L+R}(1)$. This represents the Nambu-Goldstone theorem 
[17] at  finite temperature in the model.
\section{The real-time formalism}
In the real-time formalism, the Lagrangian of the four-fermion
interactions will be
\begin{equation}
{\cal L}^R_{4F}=
\frac{G}{4}\sum_{a=1}^{2}\{[{(\bar{\psi}\psi)}^{(a)}]^2-[{(\bar{\psi}\gamma_5\psi)}^{(a)}]^2\}(-1)^{a+1}, 
\end{equation}
where $a=1$ denotes physical fields and $a=2$ ghost fields. As a result of the  
thermal  condensates  ${\langle{(\bar{\psi}\psi)}^{(a)}\rangle}_T\neq 0, (a=1,2)$,  the gap equation becomes 
\begin{equation}
1=GI, \; \; I=2N\int\frac{d^4l}{(2\pi)^4}\left[
\frac{i}{l^2-m^2+i\varepsilon}-2\pi\delta(l^2-m^2)\sin^2\theta(l^0,\mu)\right],
\end{equation}
which is actually identical to (1) obtained in the imaginary-time formalism [14]. 
\\
The propagator for the scalar bound state $\phi_S$ is now a $2\times 2$ matrix whose 
elements correspond to the four-point amputated functions $\Gamma_S^{ba}(p)$ for 
the transition from  $(\bar{\psi}\psi)^{(a)}$ to $(\bar{\psi}\psi)^{(b)} \; (a,b=1,2)$. $\Gamma_S^{ba}(p)$ obey the Schwinger-Dyson equations  
\[
\Gamma_S^{bc}(p)[\delta^{ca}-GN^{ca}(-1)^{a+1}]=
i\frac{G}{2}\delta^{ba}(-1)^{a+1},
\]
\begin{equation}
N^{ca}(p)=-\frac{i}{2}N\int \frac{d^4l}{{(2\pi)}^4}
            {\rm tr}\left[iS^{ca}(l,m)iS^{ac}(l+p,m)\right],
\end{equation}
where  $iS^{ca}(l,m)$ is the elements of the thermal matrix propagator for the 
fermions [4].
After using the gap equation (11), the solution of (12) can be 
expressed by the matrix [16]
\begin{eqnarray}
\left(\matrix{
\Gamma_S^{11}(p)& \Gamma_S^{12}(p)\cr
\Gamma_S^{21}(p)& \Gamma_S^{22}(p)\cr}\right)&=&
\frac{1}{|K(p)+H(p)-iS(p)|^2-R^2(p)} \nonumber \\
&& {} \nonumber \\
&&
\times\left(\matrix{\dfrac{-i[K^*(p)+H(p)+iS(p)]}{p^2-4m^2+i\varepsilon}
              &\dfrac{-(p^2-4m^2)R(p)}{( p^2-4m^2)^2+\varepsilon^2}\cr
               \dfrac{-(p^2-4m^2)R(p)}{( p^2-4m^2)^2+\varepsilon^2}
              &\dfrac{i[K(p)+H(p)-iS(p)]}{p^2-4m^2-i\varepsilon}\cr}
         \right),
\end{eqnarray}
where
\begin{eqnarray}
S(p)&=&4\pi^2N\int
       \frac{d^4l}{{(2\pi)}^4}\delta(l^2-m^2)\delta[(l+p)^2-m^2] \nonumber \\
    &&\times [\sin^2\theta(l^0+p^0,\mu)\cos^2\theta(l^0,\mu)+\cos^2\theta(l^0+p^0,
       \mu)\sin^2\theta(l^0, \mu)]
\end{eqnarray}
and
\begin{equation}
R(p)=2\pi^2N\int
\frac{d^4l}{{(2\pi)}^4}\delta(l^2-m^2)\delta[(l+p)^2-m^2]
\sin 2\theta(l^0,\mu)\sin 2\theta(l^0+p^0, \mu)
\end{equation}
are the terms containing the pinch singularities. $\Gamma_S^{11}(p)$ in (13)
is of the form of a causal propagator , hence if it is identified 
with the physical propagator for the scalar bound state $\phi_S$, as made in Refs. 
[16,18], then the main feature of  $\phi_S$  including its mass $2m$  could  be
shown. However, the expression of $\Gamma_S^{11}(p)$ has considerable difference 
from $\Gamma_{IF}^{\phi_S}(p)$ in (5) in the imaginary-time formalism. For finding 
a closer correspondence between the physical propagators for $\phi_S$ in the two 
formalisms, we will seek a thermal transformation matrix ${\bf \sf M}$ which can 
diagonalize the matrix propagator (13) so that
\begin{equation}
\left(\matrix{
\Gamma_S^{11}(p)& \Gamma_S^{12}(p)\cr
\Gamma_S^{21}(p)& \Gamma_S^{22}(p)\cr}\right)={\bf\sf M}^{-1}
\left(\matrix{
\Gamma_{RF}^{\phi_S}(p)& 0\cr
0& {\Gamma_{RF}^{\phi_S}}^*(p)\cr}\right){\bf\sf M}^{-1},
\end{equation}
where
\begin{equation}
{\bf\sf M}=\left(\matrix{\cosh \Theta &\sinh \Theta\cr
                \sinh \Theta & \cosh \Theta \cr}\right)
\end{equation}
and $\Gamma_{RF}^{\phi_S}(p)$ is now defined as the physical causal propagator for 
$\phi_S$.
It is seen from (13) that $\Gamma_S^{22}(p)=[\Gamma_S^{11}(p)]^*$ and
$\Gamma_S^{21}(p)= \Gamma_S^{12}(p)= {\Gamma_S^{12}(p)}^*$, thus (16) may 
be reduced to three independent algebraic equations for ${\rm Re}\Gamma_{RF}^{\phi_S}(p)$, 
${\rm Im}\Gamma_{RF}^{\phi_S}(p)$ and $\sinh \Theta$ (or $\cosh \Theta$ )
\begin{eqnarray}
{\rm Re}\Gamma_S^{11}(p)
&=&(\cosh^2\Theta +\sinh^2\Theta){\rm Re}\Gamma_{RF}^{\phi_S}(p),\;
{\rm Im}\Gamma_S^{11}(p)={\rm Im}\Gamma_{RF}^{\phi_S}(p), \;\nonumber \\
\Gamma_S^{12}(p)&=&-2 \sinh\Theta \cosh\Theta {\rm Re}\Gamma_{RF}^{\phi_S}(p).
\end{eqnarray}
Considering (13), we obtain from (18) 
\begin{equation}
\frac{\cosh^2\Theta +\sinh^2\Theta}{2\sinh\Theta\cosh\Theta}=\frac{S'(p)}{R(p)}, \ \  S'(p)=S(p)-{\rm Im}K(p),
\end{equation}
and furthermore,
\begin{equation}
\cosh\Theta=\frac{1}{\sqrt{2}}\left[\frac{S'(p)}{\sqrt{{S'}^2(p)-R^2(p)}}+1 \right]^{1/2},\ \ 
\sinh\Theta=\frac{1}{\sqrt{2}}\left[\frac{S'(p)}{\sqrt{{S'}^2(p)-R^2(p)}}-1 \right]^{1/2}.
\end{equation}
Then (18) will lead to the physical causal propagator
\begin{equation}
\Gamma_{RF}^{\phi_S}(p)=-i/(p^2-4m^2+i\varepsilon)\left[{\rm Re}K(p)+H(p)
  -i\sqrt{{S'}^2(p)-R^2(p)}\right].
\end{equation}
Based on the interactions (10) and  parallel 
derivation, we can give the pseudoscalar matrix propagator for the transition 
from $(\bar{\psi}i\gamma_5\psi)^{(a)}$ to 
$(\bar{\psi}i\gamma_5\psi)^{(b)}\; (a,b=1,2)$ with the elements
$\Gamma_P^{ba}(p) = \Gamma_S^{ba}(p)|_{m=0}$, then  diagonalize
$\Gamma_P^{ba}(p)$ by the same thermal matrix as 
${\bf\sf M}$ in (17) and obtain physical causal propagator 
$\Gamma_{RF}^{\phi_P}(p)$ for 
the pseudoscalar bound state $\phi_P=(\bar{\psi}i\gamma_5\psi)$
\begin{equation}
\Gamma_{RF}^{\phi_P}(p)=-i/(p^2+i\varepsilon)\left[{\rm Re}K(p)+H(p)
-i\sqrt{{S'}^2(p)-R^2(p)}\right].
\end{equation}
\indent The elements (20) of the matrix ${\bf\sf M}$ depend on $S(p)$, $R(p)$ and 
the imaginary part 
${\rm Im}K(p)$ of the zero-temperature loop integral and seem to have rather 
complicated expressions. However, the final result is remarkable, i.e. the matrix
${\bf\sf M}$ is identical to the thermal transformation matrix of the matrix 
propagator for an elementary scalar particle.  In fact, from the expressions (15) 
and (14) and 
the definition (4), we can write
\begin{equation}
R(p)=2\pi^2N\int\frac{d^4l}{(2\pi)^4} \delta(l^2-m^2) \delta[(l+p)^2-m^2]
\frac{\eta(l^0)\eta(l^0+p^0)}{\cosh[\beta(l^0-\mu)/2] \cosh[\beta (l^0+p^0-\mu)/2)]}
\end{equation}
and
\begin{equation}
S(p)=\cosh(\beta p^0/2)R(p)+D(p)
\end{equation}
\begin{equation}
D(p)=\frac{N}{16\pi^2}\int\frac{d^3l}{\omega_l\omega_{l+p}}
   [\delta(p^0+\omega_l+\omega_{l+p})+\delta(p^0-\omega_l-\omega_{l+p})].
\end{equation}
On the other hand, we may calculate $K(p)$ expressed by (6) in terms of the residue 
theorem and obtain
\[
K(p)=\frac{N}{16\pi^3} \int\frac{d^3l}{\omega_l\omega_{l+p}}
\left[\frac{1}{ p^0+\omega_l+\omega_{l+p}-i\varepsilon}-
\frac{1}{ p^0-\omega_l-\omega_{l+p}+i\varepsilon}\right].
\]
Hence the imaginary part of $K(p)$ 
\begin{equation}
{\rm Im}K(p)=D(p).
\end{equation}
It is seen from (26) and (25) that, ${\rm Im}K(p)\neq 0$ only if $\delta[{p^0}^2-(\omega_l+\omega_{l+p})^2]\neq 0$ or
$p^2=(\omega_l+\omega_{l+p})^2$. But the latter can be satisfied only if $p^2\geq 4m^2$. This reproduces the former conclusion that $K(p)$ is real when $p^2<4m^2$. 
The equations (24) and (26) show that we may separate the imaginary part ${\rm Im}K(p)$ 
of the zero-temperature loop integral from $S(p)$ and this fact is essential for the 
following results. Substituting (26) into (24) and considering (19), we will have
\begin{equation}
S'(p)=S(p)-{\rm Im}K(p)= \cosh(\beta p^0/2)R(p),
\end{equation}
then from (20) obtain
\begin{equation}
\cosh\Theta=[1+n(p^0)]^{1/2}, \ \ \sinh\Theta=[n(p^0)]^{1/2}, \ \ n(p^0)=1/(e^{\beta|p^0|}-1),
\end{equation}
which are precisely the elements of the thermal transformation matrix of the 
matrix propagator for an elementary scalar particle with zero chemical potential[4], 
though now we are dealing with the scalar and pseudoscalar bound states $\phi_S$ and 
$\phi_P$ composed of fermions and antifermions in the NJL model. This implies that 
scalar particles, whether elementary or composite, seem always to have the same 
thermodynamic property. 
\section{Equivalence of the two formalisms and causal, retarded and advanced propagators}
 We will prove that $\Gamma_{IF}^{\phi_S}(p)$ and $\Gamma_{IF}^{\phi_P}(p)$ 
expressed by (5) and (9) in the imaginary-time formalism are respectively the 
same as $\Gamma_{RF}^{\phi_S}(p)$ and $\Gamma_{RF}^{\phi_P}(p)$ expressed by 
(21) and (22) in the real-time formalism.  It is easy to see that, for this purpose, 
we only need prove that $K(p)+H(p)-iS^I(p)={\rm Re}K(p)+H(p)-i\sqrt{{S'}^2(p)-R^2(p)}$ or
\begin{eqnarray}
S^I(p)&=&\sqrt{{S'}^2(p)-R^2(p)}+{\rm Im}K(p) \nonumber \\
      &=&\sinh(\beta |p^0|/2)R(p)+ {\rm Im}K(p),
\end{eqnarray}
 where (27) has been used.  In fact, from (8) together with (4), (23) (25) and (26) we can obtain
\begin{eqnarray*}
S^I(p)&=&\eta(p^0)\sinh(\beta p^0/2)2\pi^2N\int\frac{d^4l}{(2\pi)^4} \delta(l^2-m^2) \delta[(l+p)^2-m^2] \\
&&\times\frac{\eta(l^0)\eta(l^0+p^0)}{\cosh[\beta(l^0-\mu)/2]\cosh[\beta(l^0+p^0-\mu)/2)]} \\
&& +\eta(p^0)2\pi^2N\int\frac{d^4l}{(2\pi)^4} \delta(l^2-m^2) \delta[(l+p)^2-m^2]
   [\eta(l^0+p^0)-\eta(l^0)] \\
&=&\eta(p^0)\sinh(\beta p^0/2)R(p) \\
 &&+\eta(p^0) \frac{N}{16\pi^2}\int\frac{d^3l}{\omega_l\omega_{l+p}}
   [-\delta(p^0+\omega_l+\omega_{l+p})+\delta(p^0-\omega_l-\omega_{l+p})] \\
&=&\sinh(\beta|p^0|/2)R(p)+ \frac{N}{16\pi^2}\int\frac{d^3l}{\omega_l\omega_{l+p}}
   [\delta(p^0+\omega_l+\omega_{l+p})+\delta(p^0-\omega_l-\omega_{l+p})] \\
&=&\sinh(\beta|p^0|/2)R(p)+{\rm Im}K(p),
\end{eqnarray*}
i.e. (29) is valid indeed. Here separation of ${\rm Im}K(p)$  from $S^I(p)$ is 
also essential.  Hence we can reach the conclusion that, in a NJL model, the four-point amputated functions corresponding to a scalar or pseudoscalar bound 
state are identical in the imaginary-time and the real-time formalism of thermal field 
theory.  This is not surprising because the calculations of four-point amputated 
functions in a NJL model can be effectively reduced to the ones of usual two-point 
functions, but with a new feature that the propagators now discussed for the bound 
states, are only subtracted four-point amputated functions, rather than the whole 
of them, because some parts of them have been subtracted through the use of the gap 
equation.  It is emphasized that the key-point of proving such equivalence and 
deriving the matrix elements (28) of ${\bf\sf M}$ lies in that one must carefully 
consider and separate the imaginary part ${\rm Im}K(p)$ of the zero-temperature loop 
integral from relevant expressions e.g. $K(p)$, $S(p)$ and $S^I(p)$, which is often 
possibly ignored in usual calculations [11]. \\
\indent Since the causal propagators for scalar or pseudoscalar bound state are 
identical in the two formalisms, we can omit the subscript $"I"$ and $"R"$ and uniquely 
express them respectively by \begin{eqnarray}
\Gamma_{F}^{\phi_S}(p)&=&\Gamma_{IF}^{\phi_S}(p)=\Gamma_{RF}^{\phi_S}(p)
                         \nonumber \\
                      &=&-i/(p^2-4m^2+i\varepsilon)[{\rm Re}K(p)+
                          H(p)-i \sinh(\beta|p^0|/2)R(p)] 
\end{eqnarray}
and
\begin{eqnarray}
\Gamma_{F}^{\phi_P}(p)&=&\Gamma_{IF}^{\phi_P}(p)=\Gamma_{RF}^{\phi_P}(p)
                       \nonumber \\
                      &=&-i/(p^2+i\varepsilon)[{\rm Re}K(p)+
                         H(p)-i \sinh(\beta|p^0|/2)R(p)]. 
\end{eqnarray}
Based on the identity of the causal propagators in the two formalisms, it is 
easy to obtain the retarded and the advanced propagator $\Gamma_r^{\phi}(p)$ and  
$\Gamma_a^{\phi}(p)$ (where $\phi$ for $\phi_S$ or $\phi_P$) which are also the same 
respectively in the two formalisms. If first in the imaginary-time formalism, then 
we may define
\begin{eqnarray}
\Gamma^{\phi}_{IF}(p)&\equiv & i \Gamma^{\phi}_I(-i\Omega_m\to p^0+i \varepsilon p^0), \nonumber \\
\Gamma^{\phi}_{Ir}(p)&\equiv & i \Gamma^{\phi}_I(-i\Omega_m\to p^0+i \varepsilon ), \nonumber \\
\Gamma^{\phi}_{Ia}(p)&\equiv & i \Gamma^{\phi}_I(-i\Omega_m\to p^0-i \varepsilon ). 
\end{eqnarray}
From (32), we can derive
\[
\Gamma^{\phi}_{IF}(p)=\theta(p^0) \Gamma^{\phi}_{Ir}(p)+ \theta(-p^0) \Gamma^{\phi}_{Ia}(p),
\]
\[
[\Gamma^{\phi}_{Ir}(p)]^*=- \Gamma^{\phi}_{Ia}(p),
\]
and furthermore,
\begin{equation}
\Gamma^{\phi}_{Ir}(p)=\theta(p^0)\Gamma^{\phi}_{IF}(p)-\theta(-p^0) [\Gamma^{\phi}_{IF}(p)]^*.
\end{equation}
Identifying $\Gamma^{\phi}_{IF}(p)$ in (33) with the common $\Gamma^{\phi}_{F}(p)$ in the two formalisms expressed by (30) and (31), we will have
\begin{eqnarray}
\Gamma^{\phi_S}_{r}(p)&=&-i/(p^2-4m^2+i\varepsilon p^0)[{\rm Re}K(p)+H(p)-i \sinh(\beta p^0/2)R(p)], \nonumber \\ 
\Gamma^{\phi_P}_{r}(p)&=&-i/(p^2+i\varepsilon p^0)[{\rm Re}K(p)+H(p)-i \sinh(\beta p^0/2)R(p)], \nonumber \\
\Gamma^{\phi}_{a}(p)&=&-[\Gamma^{\phi}_{r}(p)]^*, \ \ \phi = \phi_S \ {\rm or} \ \phi_P,
\end{eqnarray}
which represent the retarded and the advanced propagators for the bound state $\phi$.
The result (34) can also be obtained from the matrix propagator (13) in the real-time 
formalism by so called the transformation in the RA basis [10], which will be discussed 
elsewhere. 
\section{Conclusions}
We have proven that the physical causal (as well as retarded and advanced) propagators
for scalar bound states in a chiral $U_L(1)\times U_R(1)$ NJL model, defined by the 
four-point amputated functions subtracted through the use of the gap equation, are 
identical in the imaginary-time and real-time-time formalism.  This result 
convincingly shows equivalence of the two formalisms of thermal field theory in the 
NJL model. The key-points to complete the above proof lie in keeping the general form 
of the analytic continuation of the Matsubara frequency in the imaginary-time 
formalism and  separating carefully the imaginary part of the zero temperature loop 
integral from relevant expressions, and these are also certainly of crucial 
importance for general explicit demonstration of equivalence of the two formalisms, 
for instance, in the calculations to many-loop order and/or of n-point Green 
functions. In addition, we have found that, in the real-time formalism, the thermal 
transformation matrices of the matrix propagators for scalar bound states are 
precisely the one for an elementary scalar particle and this fact strongly indicates 
similarity of thermodynamic property between a composite and an elementary scalar particle. \\

\end{document}